\documentclass[fleqn,usenatbib]{mnras}

\usepackage{newtxtext,newtxmath}
\usepackage[T1]{fontenc}
\usepackage{ae,aecompl}

\usepackage{amssymb}
\usepackage{amsmath}
\usepackage{graphicx}
\usepackage{natbib}
\usepackage{color}
\usepackage{soul}

\newcommand{\He}{He\;\textsc{ii} }
\newcommand{\OIII}{[O\;\textsc{iii}] }
\newcommand{\Heline}{He\;\textsc{ii} 4686\AA\ }
\newcommand{\OIIIline}{[O\;\textsc{iii}] 5007\AA\ }

\newcommand{\SNRA}{SNR 0509-67.5}
\newcommand{\SNRB}{SNR 0505-67.9}
\newcommand{\SNRC}{SNR 0509-68.7}
\newcommand{\SNRD}{SNR 0519-69.0}

\title[Excluding SSSs as SNe Ia progenitors]{Excluding super-soft X-ray sources as progenitors for four Type Ia supernovae in the Large Magellanic Cloud}

\author[J. Kuuttila et al.]{
J. Kuuttila$^{1}$,
M. Gilfanov$^{1,2}$,
I. R. Seitenzahl$^{3,4}$,
T. E. Woods$^{5}$
and F. P. A. Vogt$^{6,7}$
\\
$^{1}$Max Planck Institute for Astrophysics, Karl-Schwarzschild-Str. 1, Garching b. M\"unchen 85741, Germany\\
$^{2}$Space Research Institute, Profsoyuznaya 84/32, 117997, Moscow, Russia\\
$^{3}$School of Science, University of New South Wales, Australian Defence Force Academy, Canberra, ACT 2600, Australia\\
$^{4}$Research School of Astronomy and Astrophysics, Australian National University, Canberra, ACT 2611, Australia\\
$^{5}$Institute for Gravitational Wave Astronomy and School of Physics and Astronomy, University of Birmingham, \\Birmingham B15 2TT, United Kingdom\\
$^{6}$European Southern Observatory, Av. Alonso de C\'ordova 3107, 763 0355 Vitacura, Santiago, Chile\\
$^{7}$ESO Fellow
}

\date{Accepted XXX. Received YYY; in original form ZZZ}

\pubyear{2018}

\begin{document}
\label{firstpage}
\pagerange{\pageref{firstpage}--\pageref{lastpage}}
\maketitle

\begin{abstract}
Type Ia supernovae are vital to our understanding of the Universe due to their use in measuring cosmological distances and their significance in enriching the interstellar medium with heavy elements. They are understood to be the thermonuclear explosions of white dwarfs, but the exact mechanism(s) leading to these explosions remains unclear. The two competing models are the single degenerate scenario, wherein a white dwarf accretes material from a companion star and explodes when it reaches the Chandrasekhar limit, and the double degenerate scenario, wherein the explosion results from a merger of two white dwarfs. 
Here we report results which rule out hot, luminous progenitors consistent with the single degenerate scenario for four young Type Ia supernova remnants in the Large Magellanic Cloud. Using the integral field spectrograph WiFeS, we have searched these remnants for relic nebulae ionized by the progenitor, which would persist for up to $\sim 10^5$ years after the explosion. We detected no such nebula around any of the remnants. By comparing our upper limits with photoionization simulations performed using Cloudy, we have placed stringent upper limits on the luminosities of the progenitors of these supernova remnants. Our results add to the growing evidence disfavouring the single degenerate scenario.
\end{abstract}


%________________________________________________________________
\section{Introduction}\label{sec:intro}

Type Ia supernovae (SNe Ia) are runaway thermonuclear explosions of carbon-oxygen white dwarfs (CO WDs) \citep[see e.g.][for reviews]{Hillebrandt13, Maoz14}.
SNe Ia can be used for distance measurements on cosmic scales due to a correlation between their peak luminosity, the rate of decline after maximum light, and the colour at maximum. This so-called Phillips relation \citep{Phillips93, Phillips99} has been used to show the acceleration of the Universe's expansion \citep{Riess98, Perlmutter99}. SNe Ia are also important to the chemical evolution of galaxies, since a typical SN Ia enriches the interstellar medium (ISM) with {$\sim$}0.7 M$_{\odot}$ of iron and a similar amount of other elements \citep{Matteucci86,Wiersma11}. 

Despite their significance to our understanding of the Universe, the formation channel for SNe Ia is still uncertain. The two leading models are the single degenerate (SD) scenario, in which a single WD reaches the critical carbon ignition density through accretion from either a main-sequence or an evolved companion star \citep{Whelan73}; and the double degenerate scenario, in which the explosion results from the merger of a binary pair of WDs \citep{Iben84}.  

In the double degenerate scenario, the progenitor system is typically too faint to be detectable with current instruments prior to the explosion. In the single degenerate scenario, however, progenitor systems should be detectable both before and, most importantly, after the explosion. In particular, if steady nuclear burning of hydrogen occurs on the surface of a WD, as expected from the single degenerate channel, then in the most efficient regime for mass accumulation the accreting WD should reveal itself as a strong super-soft X-ray source (SSS) \citep{vandenHeuvel92,Kahabka97}. 

The first of these SSSs were discovered in the Large Magellanic Cloud (LMC) using the Einstein Observatory \citep[HEAO-2,][]{Long81}. SSSs are typically luminous ($ L_{\mathrm{bol}} \gtrsim 10^{37-38}$ erg s$^{-1}$) and characterized by effective temperatures of $ T_{\mathrm{eff}} \sim  10^6$ K \citep{Greiner00}. 
Due to their high temperatures and luminosities, SSSs emit significant fluxes of UV and soft X-ray photons, which will strongly ionize any surrounding interstellar gas. This in turn will create a characteristic ionization nebula 
with a typical (``Str\"omgren'') radius of 
\begin{equation}
\mathrm{R_S} = \left( \frac{3}{4 \pi} \frac{\mathrm{\dot{N}_{ph}}}{\mathrm{n^2_{ISM}} \alpha} \right) ^{\frac{1}{3}} 
 \approx 35 \mathrm{pc} \left( \frac{\mathrm{\dot{N}_{ph}}}{10^{48} \,\mathrm{s}^{-1}} \right) ^{\frac{1}{3}} \left( \frac{\mathrm{n_{ISM}}}{1\,\mathrm{cm} ^{-3}} \right) ^{-\frac{2}{3}},
\end{equation}
where $\alpha$ is the recombination coefficient, $\mathrm{\dot{N}_{ph}}$ is the number of ionizing photons per second, and $\mathrm{n_{ISM}}$ is the number density of the ISM \citep{Rappaport94,Woods16}.
In addition to the direct emission from the central SSS, significant ionization nebulae can be also produced by the emission from the accretion disc when the accretion rate is too low for the hydrogen fusion to be ignited \citep{Woods17}. 

To date, only one such emission-line nebula has been detected, surrounding a SSS in the LMC known as CAL 83 \citep{Remillard95}. These authors, in fact, conducted also imaging observations of nine other known SSSs in the LMC and SMC, but did not detect nebulae around them. 
Later, \citet{Gruyters12} observed a part of the nebula around CAL 83 using the VLT/VIMOS and detected for the first time also the \Heline emission line. 
The high ionization potential of \He (54.4 eV) requires a hot ($\gtrsim$ 10$^{5}$ K) ionizing source, making the \He emission an easily recognizable signal of an accreting WD with steady nuclear burning on its surface. Models for SSS nebulae indeed predict that these nebulae should be bright in \Heline and \OIIIline, making them distinct from other astrophysical nebulae \citep{Rappaport94,Woods16}. For this reason, the \Heline emission line has been used to constrain the suitability of SSSs as progenitors for SNe Ia, either by examining pre-explosion archival observations of the explosion site \citep{Graur14}, or by comparing the observed total \He emission of galaxies to the expected emission from population synthesis models \citep{Woods13,Johansson14}.

If a single degenerate SN Ia progenitor spends a significant amount of time as a SSS prior to the explosion, then the ionized nebula should remain detectable after the WD is destroyed in the explosion, until the majority of the ionized gas has recombined. If the gas in the nebula is initially fully ionized (i.e. $n_e \approx n_{\mathrm{ISM}}$),
the typical hydrogen recombination time can be estimated as 
\begin{equation}
\tau _{rec} = \left( n_e \alpha _B (H^0, T \approx 10^4 K ) \right) ^{-1} \approx 10^5 \times \left( \frac{n_{\mathrm{ISM}}}{1 \mathrm{cm}^{-3}} \right) ^{-1} \mathrm{years}, 
\end{equation}
where $\alpha _B (H^0, T)$ is the Case B recombination coefficient \citep{Woods16, Woods17}. For helium the corresponding time-scale is $\approx 7 \times 10^4$ years \citep[see e.g.][]{Pequignot91}.
Searching for these relic nebulae around young supernova remnants and determining the ionization state of the surrounding gas can thus be used effectively to constrain the properties of the progenitor. 

Previous efforts in determining the ionization state of the gas around SNRs have focused on the forward shocks. Many shock fronts around SNe Ia remnants are so-called Balmer-dominated shock fronts, where the optical emission is dominated by both broad and narrow Balmer line emission. This emission is understood to be the result of the shock interacting with the surrounding neutral hydrogen, and can thus be used in estimating the ionized fraction of hydrogen \citep{Ghavamian00,Ghavamian01,Ghavamian03}. This method has recently been used to place stringent upper limits on the luminosity of the progenitors for SNe Ia remnants in the Galaxy and LMC \citep{Woods17,Woods18}. Any method used to determine the ionization state of the gas is very sensitive to the density of the gas, but the expanding shocks can also be used to determine the density of the surrounding gas \citep{Badenes07,Yamaguchi14}.

Here we report a different method for constraining the nature of SNe Ia progenitors. We have searched directly for the relic ionization nebulae around four known SNe Ia remnants in the LMC (see Table~\ref{table:sources} for list of sources). 
Using integral field spectroscopy, we searched for and did not find any \Heline emission ahead of the forward shocks. With comparison to numerical simulations performed with the photoionization code Cloudy, we placed upper limits on the luminosities as a function of the effective temperatures of the progenitors that created the observed four LMC remnants. Using these upper limits, we excluded the presence of accreting nuclear burning WDs at the sites of the remnants during the last $\sim 10^5$ years before the explosions. 

This paper is organized as follows: In Sec.~\ref{sec:obs} we describe our observations and the data reduction procedure. 
In Sec.~\ref{sec:methods} we describe the methods used in this paper, specifically the spectral extraction (Sec.~\ref{sec:spectra}) and the Cloudy simulations (Sec.~\ref{sec:sim}). In Sec.~\ref{sec:results} we describe our results and then discuss them and the possible implications in Sec.~\ref{sec:disc}.

\section{Observations}\label{sec:obs}
We have observed the four LMC SN Ia remnants %listed in Table~\ref{table:sources}
\SNRA, \SNRB, \SNRC, and \SNRD\
with the Wide Field Spectrograph (WiFeS) mounted on the Nasmyth A focus of the Australian National University 2.3\,m telescope at the Siding Spring Observatory \citep{Dopita07,Dopita10}. \SNRB, \SNRC, and \SNRD\ were observed on the nights of 2014 December 18--20 (P.I.: Seitenzahl; Proposal ID: 4140118) and \SNRA\ was observed on 2015 December 13 (P.I.: Seitenzahl; Proposal ID: 4150145). Here we provide only a short summary of the data reduction method, which is also described in detail by \citet{Dopita16} and \citet{Ghavamian17}.

The observations were performed in the `binned mode', which provided us a field of view of 25 $\times$ 35 spatial pixels (or spaxels), each of them $1'' \times 1''$ in angular size. The instrument is a double-beam spectrograph providing simultaneous and independent channels for both the blue and red wavelength ranges. 
We used the B3000 and R7000 gratings, providing a spectral resolution of R = 3000 ($\Delta v \approx$ 100 km s$^{-1}$) in the blue wavelength range (3500--5700 \AA ) and R = 7000 ($\Delta v \approx$ 45 km s$^{-1}$) in the red (5300--7000 \AA ). 

\SNRA, \SNRC, and \SNRD\ were observed in a mosaic of two overlapping fields, and \SNRB\ was observed with ten fields in order to cover the whole remnant. Each field was observed in $2 \times 1800$s exposures, with $2 \times 900$s blank sky exposures, which were subtracted from the two co-added frames for each field. 

The data were reduced with the \textsc{PYWIFES} v0.7.3 pipeline \citep{Childress14ascl,Childress14}, which provided us a wavelength calibrated, sensitivity corrected, and photometrically calibrated data cube. 
The final mosaics were then combined from the individually reduced cubes, with the respective alignment of each field in the mosaic derived by comparing the reconstructed continuum frames from the red cubes with the Digitized Sky Survey 2 red band image of the area. 
The final mosaic for \SNRB \, has dimensions of $94'' \times 96''$, and for the three other sources the dimensions are $40'' \times 36''$, which correspond to fields of $22.8 \times 23.3$ pc and $9.7 \times 8.7$ pc, respectively, assuming a distance of 50 kpc to the LMC.

\begin{table*}
\caption{List of observed sources with relevant properties. }
\begin{tabular}{lccccc}
\multicolumn{1}{c}{Source}&\multicolumn{1}{c}{Size (pc)}&\multicolumn{1}{c}{Age (yrs)}&\multicolumn{1}{c}{$n_0$ (cm$^{-3}$)}&\multicolumn{1}{c}{N$_{\mathrm{H}}$ (10$^{21}$ cm$^{-2}$) $^a$}&\multicolumn{1}{c}{References}\\ \hline\hline 
SNR 0509-67.5 & 4 & 400 $\pm$ 120 & 0.4--0.6 & 1.64 $\pm$ 0.07 & \citet{Rest05, Kosenko08} \\
SNR 0505-67.9$^1$ & 7--9 & $\sim$ 4700 & 0.5--1.5 & 0.28 $\pm$ 0.001 & \citet{Hughes98,Ghavamian03} \\
SNR 0509-68.7$^2$ & 4 & 685 $\pm$ 20 & 1--2.5 & 3.09 $^{+0.20} _{-0.11}$ & \citet{vanderHeyden02,Williams14} \\
SNR 0519-69.0 & 4 & 680 $\pm$ 200 & 2.4 $\pm$ 0.2 & 0.96 $\pm$ 0.04 & \citet{Rest05,Kosenko10} \\ 
 \hline\hline
 $^1$\footnotesize{ DEM L71}\\
 $^2$\footnotesize{ N103B}\\
 $^a$\footnotesize{ \citet{Maggi16}}
\end{tabular}
\label{table:sources}
\end{table*}

\section{Methods}\label{sec:methods}

\subsection{Spectra of the remnants}\label{sec:spectra}

\begin{figure*}%[!ht]
	\centering
    \includegraphics[width=1\textwidth]{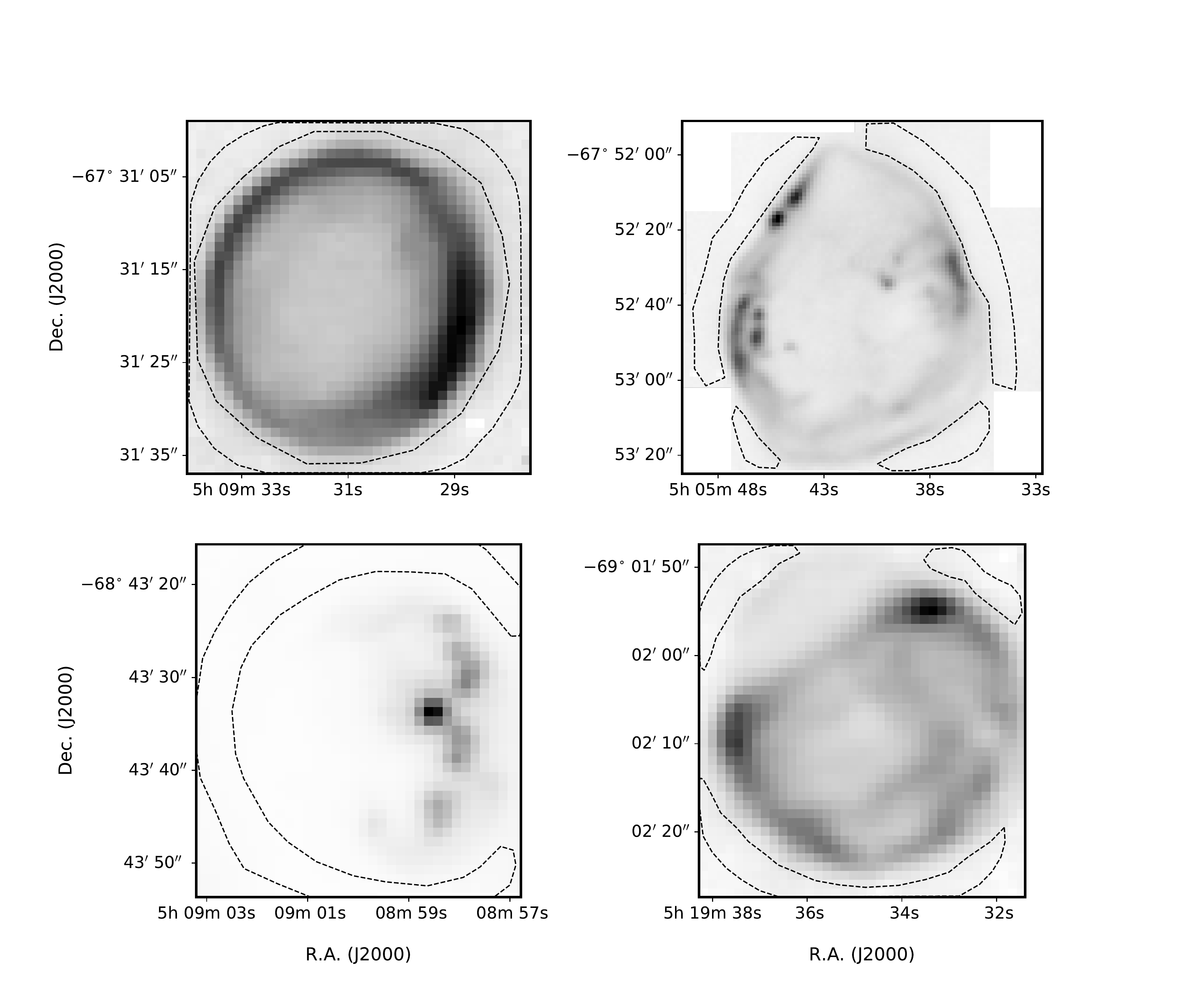}
    \caption{All supernova remnants in H$\alpha$ with WiFeS. Top row from left to right: \SNRA\ and \SNRB ; bottom row from left to right: \SNRC\ and \SNRD . The spectral extraction areas are outlined with the black dashed lines.}
\label{fig:remnantArea}
\end{figure*}

In order to study the properties of the possible nebulae around the observed supernova remnants with as high sensitivity as possible, we extracted spectra from large areas surrounding each remnant. Since the expected brightness of an emission line decreases as a function of distance (see Fig.~\ref{fig:SB_examples}), we used the area between the outer edge of the forward shock ($\sim 4$ pc from the centre) and a distance of about 5 pc from the approximate geometrical centre of each source for \SNRA, \SNRC, and \SNRD; for \SNRB~ the corresponding values are 7--9 pc and 10 pc. We avoided any areas with residuals from foreground star subtraction. As an illustrative example, see Fig.~\ref{fig:remnantArea}, where we show the \SNRD\ remnant with the spectral extraction area marked. 

\begin{figure}%[!ht]
	\centering
    \includegraphics[width=1\columnwidth]{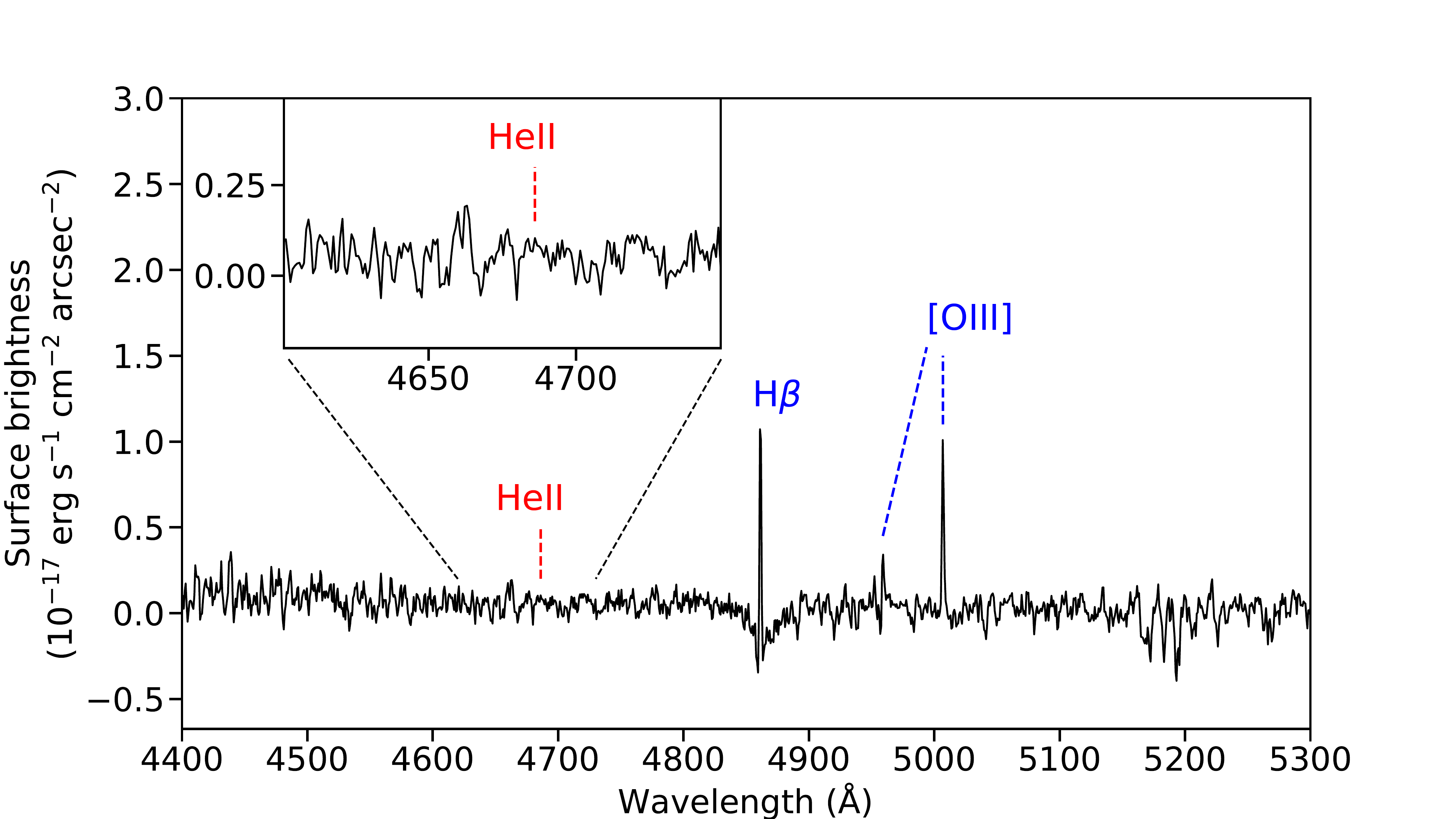}
    \caption{An example spectrum of the interstellar gas ahead of the forward shock around \SNRD . On the y-axis is the mean surface brightness in units 10$^{-17}$ erg s$^{-1}$ cm$^{-2}$ arcsec$^{-2}$ and on the x-axis is the wavelength in units of \AA . The inset figure shows the spectrum in more detail around the 4686 \AA\ wavelength. In blue are marked the brightest emission lines H$\beta$ and \OIII, and the red dashed lines indicate the 4686\AA\ wavelength. }
    \label{fig:example_spectrum}
\end{figure}

For each source, the spectra were averaged over the specified area, corrected for the average redshift of 277.5 km s$^{-1}$, which was measured from the H$\beta$ and \OIII 4959, 5007 \AA\ emission lines 
\citep[LMC peculiar velocity is 262.2 km s$^{-1}$;][]{McConnachie12}, 
and dereddened using the average LMC extinction curves of \citet{Weingartner01} with a carbon abundance $b_c = 2 \times 10^{-5}$ and using the H column densities for each source listed in table~\ref{table:sources}. An example spectrum is shown in Fig.~\ref{fig:example_spectrum}.

No \Heline emission was detected in any source before the forward shock, as is evident in Fig.~\ref{fig:example_spectrum}. We used this lack of noticeable line emission to derive upper limits for the \Heline flux by estimating the maximum level of emission which would be inseparable from the noise. 
In order to determine the noise level, we calculated the standard deviation of the observed flux within a wavelength window with an approximate width of 200 \AA , positioned around 4686 \AA\ so that no bright emission lines, namely H$\beta$, were inside the window. 

Then, because the nebular \He emission is expected to be narrower than the instrumental resolution and would thus be spectrally unresolved \citep{Gruyters12}, 
we assumed a gaussian line with a fixed width corresponding to the instrumental resolution of B3000 ($\sim 100$ km s$^{-1}$), and using a chi-squared test we calculated the minimum amplitude of a line, which would be statistically separable from the estimated noise level. The line was taken to be distinct from the noise, when adding the line on the spectrum increased the $\chi ^2$ value by 9, corresponding to $3 \sigma$, or 99.7 \% confidence. The flux of such a gaussian line was then taken to be the upper limit of the possible \Heline line flux and compared against the simulations described in Sec.~\ref{sec:sim}.

\subsection{Simulations}\label{sec:sim}
We computed a grid of numerical photoionization models with Cloudy\footnote{www.nublado.org} \citep[v17.01;][]{Ferland13}. We assumed a spherically symmetric and static configuration with the central ionizing source emitting a blackbody spectrum, which provides a reasonable approximation of the ionizing emission of nuclear-burning WDs, except far into the Wien tail \citep{Chen15, Woods16}. The effective temperature of the ionizing radiation was varied from $10^4$ to $10^7$ K and the bolometric luminosity from $10^{35}$ to $10^{38}$ erg s$^{-1}$ with logarithmically evenly spaced steps. In light of the pre-shock densities of the remnants shown in Table~\ref{table:sources}, the density of the ambient gas was kept fixed at either 0.5, 1, or 2.4 cm$^{-3}$, while dust was neglected. The metallicity of the gas was set to $Z = 0.3 \; Z _{\odot}$, where $Z _{\odot}$ is the solar metallicity, based on the average results of several studies on the metallicity of the ISM around many LMC SNRs, including for example \SNRB\ and \SNRD\ \citep{Hughes98,Maggi16,Schenck16}. 

The calculations were performed in three different ways with regard to the gas temperature: in the first case, the ambient gas temperature was calculated self-consistently and the calculations were stopped when the gas temperature dropped below 3000\,K. 
While this is an idealized assumption, this case offers a possibility to study as an example an isolated situation, where the only source of energy is the central ionizing source. In reality, there is a diffuse emission field originating from stars and other sources in addition to the central ionizing source.
Depending on the strength of the diffuse emission field, and properties of the gas, such as density, the ISM has been historically classified roughly into three different phases: a hot and very low density phase ($n \sim 10^{-2.5}$ cm$^{-3}$, $T \sim 10^6$ K), a warm low density phase ($n \sim 10^{-0.5}$ cm$^{-3}$, $T \sim 10^4$ K), and a cold dense phase ($n \sim 10^{1.5}$ cm$^{-3}$, $T \sim 10^2$ K) \citep{McKee77}. 
In the hot phase the gas is already ionized and any possible SSS would not then change the ionization state of the ISM. In the cold phase the central SSS would be the main source of energy and an ionization nebula would be clearly detectable. This phase corresponds mostly to the self-consistent temperature calculations. 
However, the estimated gas density limits of the SNRs studied here point mostly to the warm phase. Thus, to include the contribution from the diffuse emission in our simulations, we ran the calculations with a fixed gas temperature in addition to calculating it self-consistently. Although the relatively low temperatures of the warm low density phase are not expected to contribute significantly to the ionization of He$^{+}$ due to its high ionization potential (54.4 eV), we ran the calculations with the temperature set to either 5000 K or 10000 K in order to test the effect of the gas temperature on the \Heline emission.

\begin{figure}%[!ht]
	\centering
    \includegraphics[width=1\columnwidth]{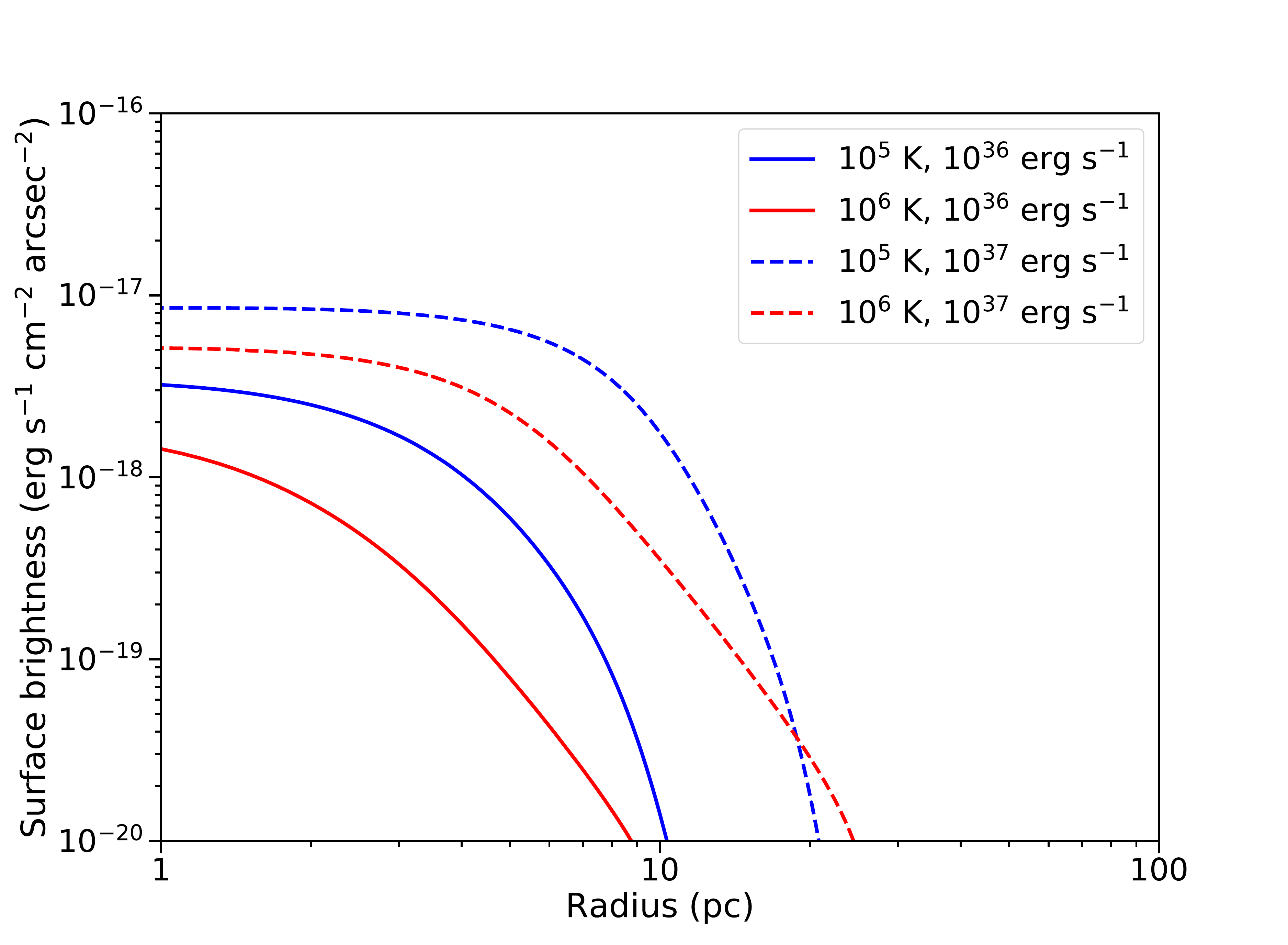}
    \caption{Surface brightness (in units of erg s$^{-1}$ cm$^{-2}$ arcsec$^{-2}$) profiles for \Heline as a function of radius (in parsecs). Blue and red lines have blackbody temperatures of 10$^5$ and 10$^6$ K, and the solid and dashed lines have luminosities of 10$^{36}$ and 10$^{37}$ erg s$^{-1}$, respectively. The density of the gas was set to 1 cm$^{-3}$ and the gas temperature was calculated self-consistently.}
\label{fig:SB_examples}
\end{figure}

From the Cloudy simulations we get the volume emissivity $\epsilon _i (r)$ of a line i as a function of the distance $r$ from the ionizing source. This can be used to find the surface brightness of a line i:
\begin{equation}
\mathrm{SB_i}(r) = \int _l \frac{\epsilon _i (r)}{4 \pi} \mathrm{d}l,
\end{equation}
where we have integrated along the line of sight \textit{l} through the emission nebula. Examples of the \Heline surface brightness as a function of radius for the central source temperatures of 10$^{5-6}$ K and luminosities 10$^{36-37}$ erg s$^{-1}$ are shown in Fig.~\ref{fig:SB_examples}. 

From these surface brightness profiles, we calculated the average surface brightness of the \Heline emission line in the data extraction range of 4--5 pc (or 7--10 pc for \SNRB) for each point in the temperature--luminosity grids. Then, comparing the upper limits acquired from the WiFeS observations to the grids of simulated brightnesses, we can constrain the luminosity as a function of the assumed emission temperature of the central ionizing sources.

\section{Results}\label{sec:results}

The $3\sigma$ upper limits for the surface brightness of the \Heline emission line acquired for each source with the method explained in Sec.~\ref{sec:spectra} are shown in Table~\ref{table:SB_results}. No noticeable \He emission was detected ahead of the forward shock in any source and the derived upper limits are within a factor of two from each other. How well the progenitor properties can be constrained, depends, however, on the size of the remnant and the density of the surrounding gas. \SNRB\ is much older and thus much larger than the other three SNRs. As shown in Fig.~\ref{fig:SB_examples}, the expected surface brightness decreases with the distance from the ionizing source, making it harder to constrain the progenitor luminosities of older and larger SNRs.

\begin{table}%[!ht]
\caption{$3 \sigma$ upper limits on the \Heline surface brightness for each source.}
\begin{tabular}{lc}
\multicolumn{1}{c}{Source}&\multicolumn{1}{c}{Surface brightness}\\
\multicolumn{1}{c}{}&\multicolumn{1}{c}{($\times \, 10^{-19}$ erg s$^{-1}$ cm$^{-2}$ arcsec$^{-2}$) }\\\hline\hline 
SNR 0509-67.5 & 4.2 \\ %2.3
SNR 0505-67.9 & 4.7 \\ %2.4
SNR 0509-68.7 & 5.7 \\ %3.8
SNR 0519-69.0 & 5.3 \\ %2.9
 \hline\hline
\end{tabular}
\label{table:SB_results}
\end{table}

\begin{figure}%[!ht]
	\centering
    \includegraphics[width=1\columnwidth]{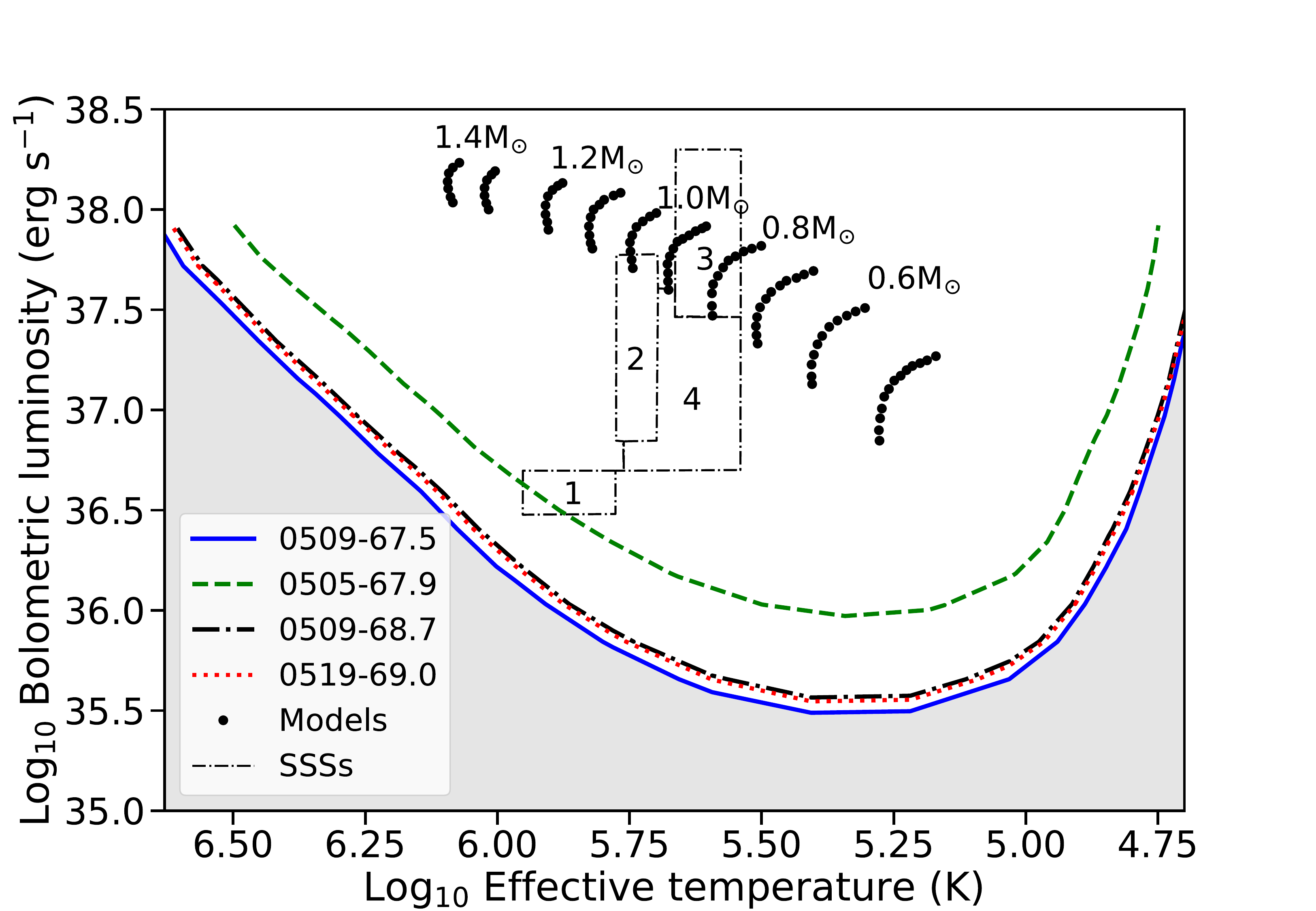}
    \caption{$3\sigma$ upper limits on the bolometric luminosity as a function of the assumed emission colour temperature for the progenitors of the four SNRs studied here. The blue, green, black, and red lines show the upper limits for \SNRA , \SNRB , \SNRC , and \SNRD , respectively. 
In all cases in this figure, the ambient gas density is set at 1 cm$^{-3}$, the temperature is calculated self-consistently, and the calculations terminated when the temperature dropped below 3000 K. For comparison, the black dotted lines show the accreting nuclear-burning WD models of \citet{Wolf13} with the mass increasing from 0.51 M$_{\odot}$ on the left to 1.4 M$_{\odot}$ on the right. For ease of reading, only every second model is labelled. The black dash-dotted boxes represent the parameter ranges of four well-known SSSs: 1. CAL87; 2. 1E 0035.4-7230; 3. RX J0513.9-6951; and 4. CAL 83 \citep{Greiner00}. }
\label{fig:grid_allSources}
\end{figure}

To transform the surface brightness upper limits to limits on the progenitor luminosities, we compared the results to the Cloudy simulations, as explained in Sec.~\ref{sec:sim}. The upper limits on the bolometric luminosity as a function of the assumed emission colour temperature for each source are shown in Fig.~\ref{fig:grid_allSources}. 
In this figure, the parameter space above each line is ruled out, and the area below is unconstrained. Here the temperature is calculated self-consistently and the gas density is set to 1 cm$^{-3}$. The limits of the three young and small remnants are all almost the same. \SNRB\ deviates from the others mostly because of its greater size; the surface brightness around \SNRB\ is studied at a distance of $\approx$ 8 pc, which together with the expected surface brightness profiles of Fig.~\ref{fig:SB_examples} results in lower constraints. 

For comparison, in Fig.~\ref{fig:grid_allSources} are shown also the temperatures and luminosities of the accreting nuclear-burning white dwarf models of \citet{Wolf13}, with the white dwarf mass ranging from 0.51 M$_{\odot}$ to 1.4 M$_{\odot}$. All of these models lie well above the derived upper limits for all the SNRs studied in this paper. 
In Fig.~\ref{fig:grid_allSources} are also shown the parameter ranges for four well-known super-soft X-ray sources located in the Magellanic clouds: 1. CAL\,87 (LMC); 2. 1E\,0035.4-7230 (SMC); 3. RX\,J0513.9-6951 (LMC); and 4. CAL\,83 (LMC) \citep{Greiner00}. All of these four SSSs lie in the ruled-out regions of the three young sources, with the latter three SSSs having similar temperatures and luminosities as the nuclear-burning WD models. The upper limit of the largest remnant, \SNRB , overlaps with the parameter range of CAL 87, but one should note that CAL 87 (number 1 in Fig.~\ref{fig:grid_allSources}), which has the lowest claimed luminosity of the four, is viewed almost edge-on, meaning that its unobscured luminosity is likely much higher \citep{Ness13}.

\begin{figure}%[!ht]
	\centering
    \includegraphics[width=1\columnwidth]{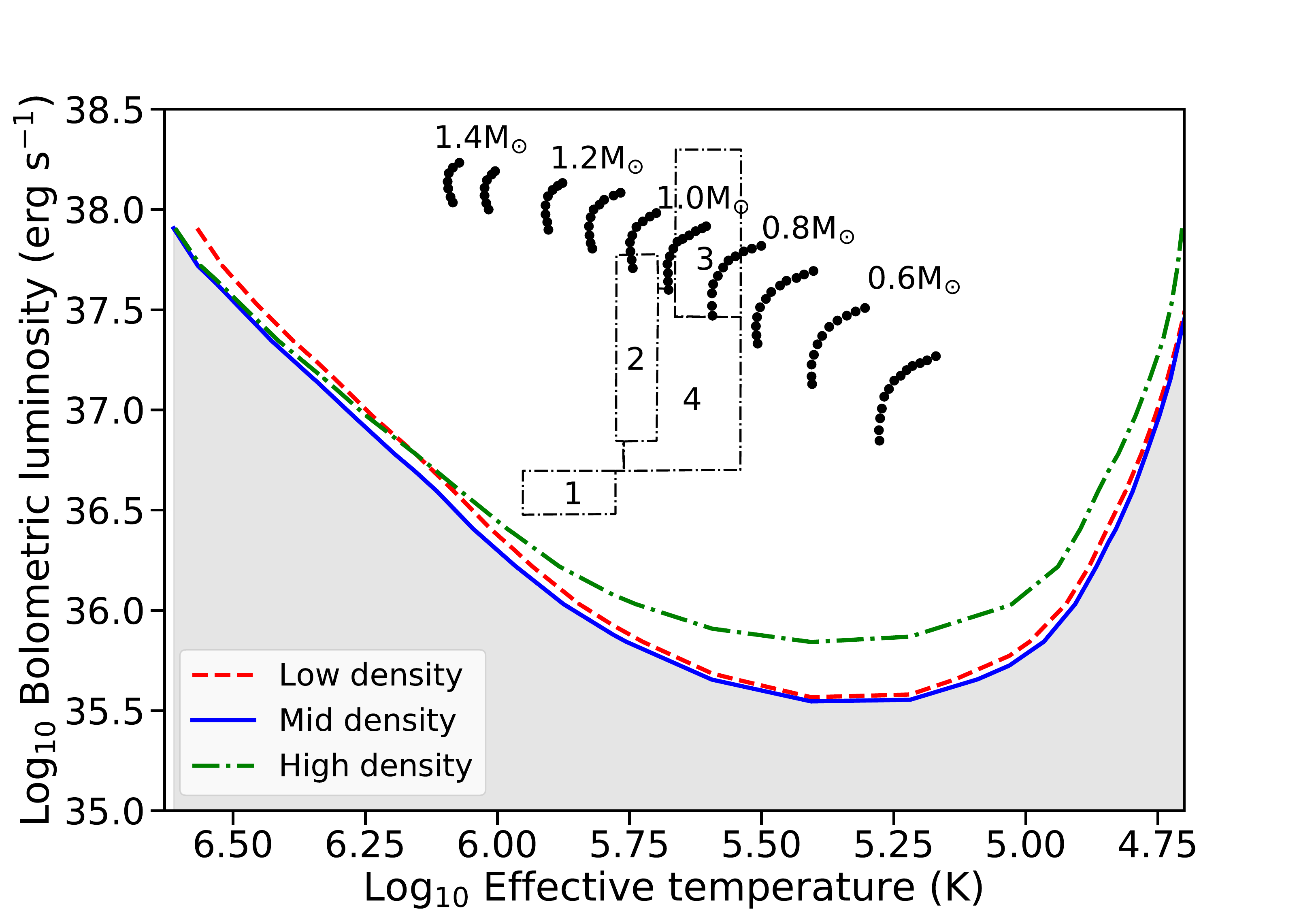}
    \caption{$3\sigma$ upper limits on the bolometric luminosity of the progenitor of \SNRD\ with different densities. The densities 0.5, 1, and 2.4 cm$^{-3}$ are shown in red dashed, blue solid, and green dot--dashed lines, respectively. Also shown are the nuclear burning WD models and SSSs, as in Fig.~\ref{fig:grid_allSources}. }
\label{fig:density_grid}
\end{figure}

In Fig.~\ref{fig:grid_allSources}, the results for each source are shown with the gas density of the simulations set to 1 cm$^{-3}$, but as mentioned before, the results are affected by the assumed gas density of the simulations. To test this effect, we ran the simulations also with the gas density set either to 0.5 or 2.4 cm$^{-3}$, which correspond to the upper limits of the density around \SNRA\ and \SNRD\ (see Table.~\ref{table:sources}), respectively. The effect of the density on the results is shown in Fig.~\ref{fig:density_grid}. As is evident from this figure, the highest density provides the least constraining limits, while the low and mid density limits differ only slightly from each other, with the mid density limits being the most constraining. 

\begin{figure}%[!ht]
	\centering
    \includegraphics[width=1\columnwidth]{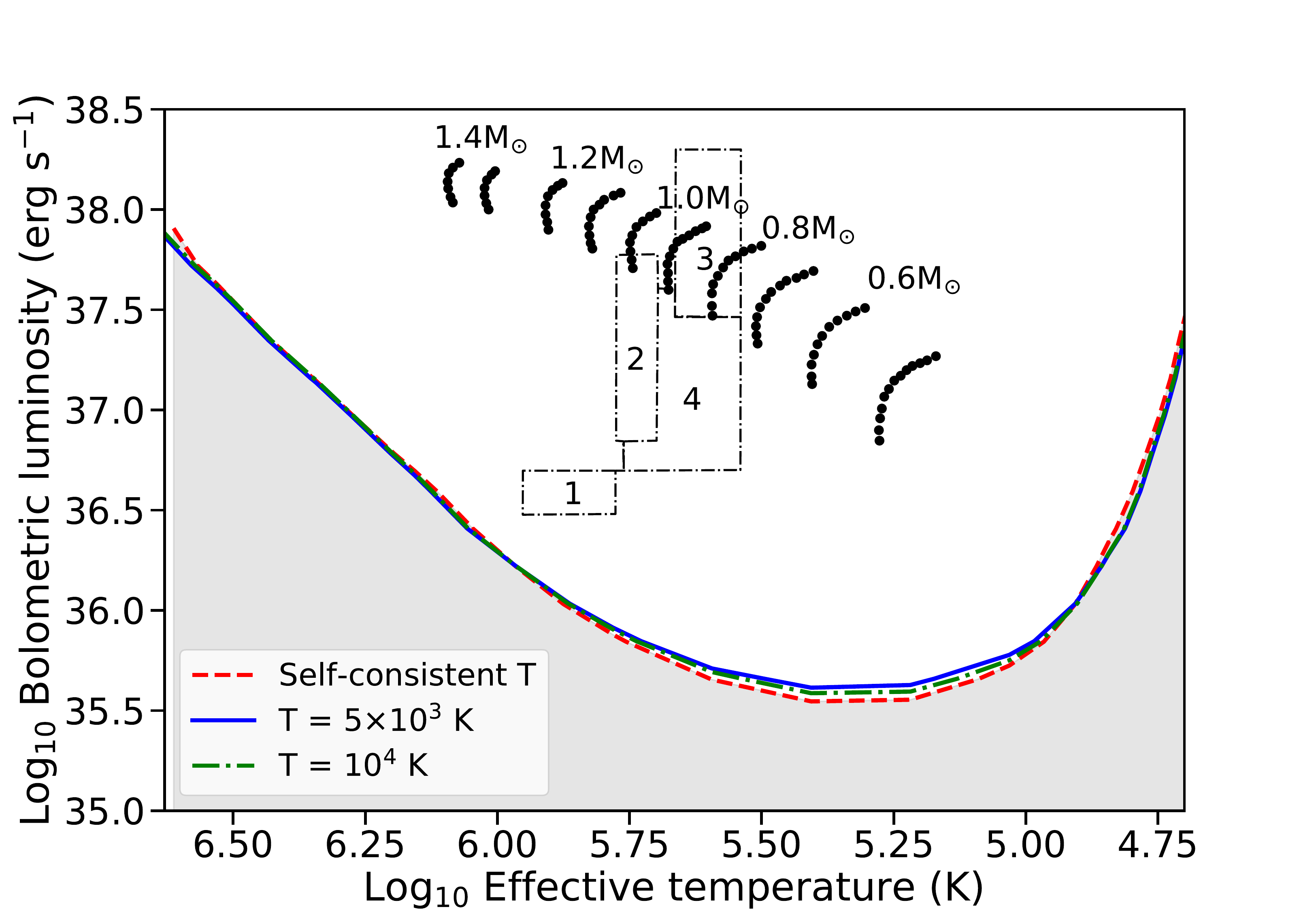}
    \caption{$3\sigma$ upper limits on the bolometric luminosity for \SNRD\ with different electron temperatures. The temperatures are either calculated self-consistently (red dashed line) or fixed at 5000 (blue solid) or 10000 K (green dot--dashed). The density is set to 1 cm$^{-3}$ in all cases. Also shown are the nuclear burning WD models and SSSs, as in Fig.~\ref{fig:grid_allSources}. }
\label{fig:temp_grid}
\end{figure}

In addition to the density, we tested how the assumed temperature of the gas affects the results. This effect is demonstrated in Fig.~\ref{fig:temp_grid}, where we show the upper limits with the temperature either calculated self-consistently, fixed at 5000 K, or fixed at 10000 K. From this figure one can see that the chosen simulation temperature affects the results only very little. The reason for this is the high ionization potential of He \textsc{II} (54.4 eV), which requires much higher energies than available in a typical warm interstellar medium. 

\begin{figure}%[!ht]
	\centering
    \includegraphics[width=1\columnwidth]{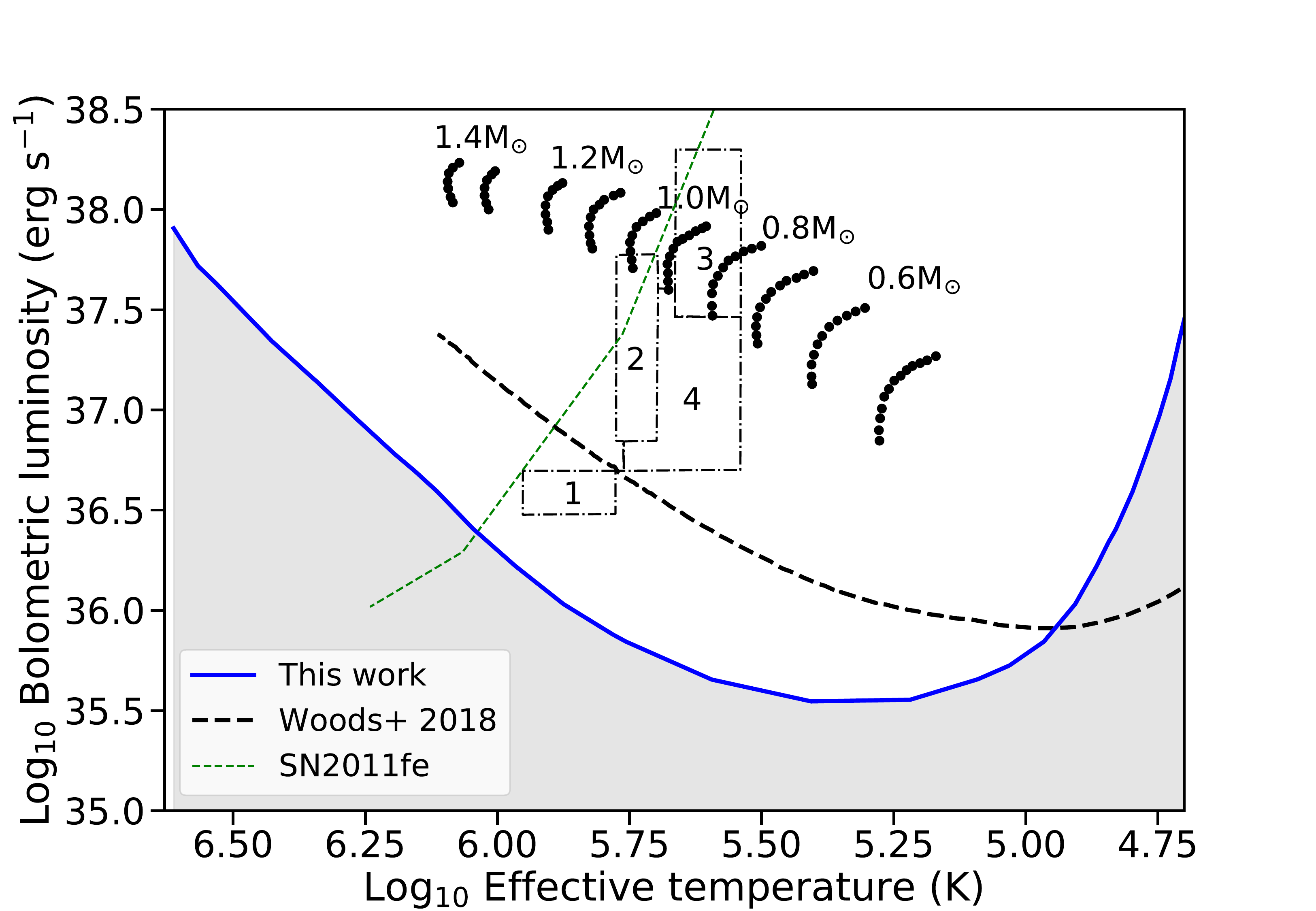}
    \caption{$3\sigma$ upper limits on the bolometric luminosity as a function of the emission colour temperature for the progenitor of \SNRD . In this figure, the ambient gas density is set at 1 cm$^{-3}$, the temperature is calculated self-consistently and the calculations terminated when the temperature dropped below 3000 K. The solid blue line shows the upper limits derived in this paper, and for comparison the black dashed line shows the upper limit for \SNRD\ derived by \citet{Woods18} using the Balmer-dominated shocks. The green dashed line shows the upper limit from pre-explosion archival X-ray data for SN 2011fe \citep{Nielsen12}. Also shown are the nuclear burning WD models and SSSs, as in Fig.~\ref{fig:grid_allSources}.}
\label{fig:comparison_grid}
\end{figure}

The upper limits on the bolometric luminosity as a function of the assumed emission colour temperature for the progenitor of \SNRD\ are also shown in Fig.~\ref{fig:comparison_grid}. Based on the analysis presented in this paper, the parameter space above the blue line is ruled out. For comparison, the upper limits for the same source derived by \citet{Woods18} are shown in the same figure with a black dashed line. For effective temperatures higher than $\sim 10^5$ K, our analysis provides significantly tighter constraints on the bolometric luminosity than that of \citet{Woods18}, who derived the limits using the Balmer-dominated forward shocks of the supernova remnant \citep[see also e.g.][]{Ghavamian03,Woods17}. On the other hand, for temperatures lower than $\sim 10^5$ K, the work of \citet{Woods18} provides lower upper limits on the luminosity than ours, because
the incident radiation field does not possess significant amount of photons with sufficient energies to ionize He$^+$ ions, causing this regime to be poorly constrained by our work, while \citet{Woods18} rely on the ionization of hydrogen, which requires considerably lower photon energies. In Fig.~\ref{fig:comparison_grid} is also shown for comparison the upper limits derived from pre-explosion archival \textit{Chandra} X-ray data for SN2011fe, which has the lowest upper limits of the ten SNe Ia studied by \citet{Nielsen12}. The upper limits for SN2011fe are slightly lower than our results for \SNRD\ in high temperatures ($\gtrsim 10^6$ K), for the part that there exists data for SN2011fe. In the high temperature regime our results become less constraining, because increasing the photon energies leads to less efficient ionizing of the ambient gas, which is due to the ionizing cross section of a hydrogen-like ion decreasing as a function of energy, approximately as $\sigma \propto E^{-3}$ \citep{Hummer63}.

\section{Discussion}\label{sec:disc}

The super-soft X-ray sources have long been suggested as possible progenitors for Type Ia supernovae. However, recent studies have constrained their viability as a progenitor channel both for large populations \citep{diStefano10,Gilfanov10,Woods13,Johansson14,Woods16} and individual supernova remnants \citep{Nielsen12,Woods17,Woods18,GW18}. In this paper, we have presented a novel method for constraining supernova progenitor properties, and using this method, we have strongly disfavoured the super-soft progenitor channel for four Type Ia supernova remnants in the Large Magellanic Cloud.

With this method we have focused only on the \Heline emission line, because its high ionization potential makes it a very distinct signature of a conventional super-soft source. However, there are also other emission lines, e.g. H$\beta$ and \OIII 5007\AA , present in the spectra of the ISM around the SNRs, as is evident from Fig.~\ref{fig:example_spectrum}. These emission lines, while expected to be bright in a SSS nebula \citep{Rappaport94}, are present also in a typical warm ISM in the LMC \citep[e.g.][]{Pellegrini12} and thus with these lines one encounters the problem of disentangling the ionization caused by the possible progenitor from the contributions of other sources, such as the diffuse background and the shock emission \citep{Smith94,Ghavamian00}. For example, in the case of the most luminous allowed (by the \He analysis) source with a temperature of 10$^5$ K, the predicted H$\beta$ emission line brightness is a factor of 5 lower and \OIIIline brightness is 10 times lower than observed around \SNRD .

Our results add to the growing body of evidence supporting the double degenerate scenario as a progenitor channel for these remnants. For \SNRD\ \citet{Edwards12} ruled out all post-main-sequence stars as possible surviving ex-companions and thus claim that among the published single-degenerate models, only the super-soft X-ray source model is a possibility for this remnant.
In addition, \SNRD\ has a tilted axisymmetric morphology and high oxygen abundance, which points to an oxygen-rich merger \citep{Kosenko10,Kosenko15}. Taking these together with our results, which rule out a SSS as a plausible progenitor, it seems clear that the only viable origin of \SNRD\ was the merger of two white dwarfs. Similarly for \SNRA\, \citet{Schaefer12} ruled out all possible surviving companion stars in the centre of the remnant, and thus ruled out all single degenerate scenarios as a progenitor channel for this remnant, which is in good agreement with our results.

Our results disfavour SSSs as possible progenitors, but we made some simplifications along the way, which should be considered in detail.
Firstly, we assumed that the luminosity remained constant throughout stable accretion and nuclear burning, although in reality these sources exhibit complex variability. However, for variable sources, the parameter of interest is the time-averaged luminosity, which determines the average ionization state of the gas, and thus given a sufficiently long time-scale, the system can be well approximated with a constant luminosity case \citep{Chiang96,Woods17}. The detailed structure of ionization nebulae may change based on the behaviour of the central source, for example in the case of nova outbursts. Such cases, and the time variability of the source and nebulae, will be addressed in future studies.

Secondly, the calculations were carried out in steady-state, i.e. assuming an equilibrium state between ionization and recombination, where the central source continues to supply the nebula with ionizing photons.
This is obviously not the case for SNRs, where the possible central ionizing source has exploded and the emission has ceased. Nevertheless, this is a reasonable assumption in the case of young SNRs, where the age (< 1000 yrs) is much smaller than the typical recombination time-scale of the ISM ($\sim$ 10$^5$ yrs). 
This argument raises the question, however, of whether there could be a long delay between the explosion and the ionizing phase. This can be achieved with spin-up/spin-down models \citep{Justham11}, where the accreting WD is spun up because of the accreted angular momentum. Because of the high spin rates, the mass of the WD can increase beyond the critical mass, and only after accretion has ceased and the spin rate of the WD has decreased can the WD explode as a supernova. If the spin-down time is longer than the recombination time, this model can produce super-Chandrasekhar single-degenerate explosions surrounded by neutral gas. In addition, by the time of the explosion, the donor star may have exhausted its stellar envelope and become a WD, rendering it difficult to detect in post-explosion companion searches \citep{diStefano11}. In fact, such super-Chandrasekhar explosions would be preferentially overluminous, ``1991T-like'' events \citep{Fisher99}. This is thought to be the case for \SNRA , which \citet{Rest08} showed to be a 1991T-like event using its light echoes, a result at which \citet{Badenes08} also arrived independently, using the remnant dynamics and X-ray spectroscopy. Super-Chandrasekhar-mass explosions, however, can also result from double-degenerate mergers, which lack the issues facing spin-up/spin-down models, such as the scarcity of observed rapidly-spinning WDs \citep{diStefano11,Maoz14}.

Thirdly, in the analysis presented here, we have considered only unobscured sources, where all the emitted radiation contributes to the ionization of the surrounding gas. The emission could, however, be obscured by a fast-moving and optically-thick stellar wind, if the WD were accreting at higher rates than the steady nuclear-burning regime \citep{Hachisu96,Wolf13}. If the wind mass-loss rate were high enough to obscure the central source, however, the wind should excavate a large ($\gtrsim $ 10 pc) low-density cavity around the progenitor, which should be easily distinguished from the undisturbed ISM. Such large cavities are incompatible with the remnants' dynamics for the three young supernova remnants studied here \citep{Badenes07}, and the densities (see Table~\ref{table:sources}) and evolution of the remnants are consistent with expansion into a uniform and undisturbed ISM \citep{Maggi16}. In addition to a wind from the accreting WD, a slow and dense wind from a giant companion star may obscure the ionizing radiation, if the mass-loss rate is $\gtrsim 10^{-6}$ M$_{\odot}$ yr$^{-1}$ \citep{Nielsen15}. 
However, such a scenario is disfavoured for SNe Ia progenitors, given the strong constrains on circumstellar interactions both from radio \citep{Chomiuk12,Chomiuk16} and X-ray observations \citep{Margutti12,Margutti14}, 
and the lack of detected giant companions \citep{Edwards12,Schaefer12,Olling15}.

Therefore, we may conclude that none of the progenitors of the Magellanic supernova remnants considered here were super-soft X-ray sources for a significant fraction of the last 100,000 years preceding their detonation. Future spectroscopic observations can extend these limits to all nearby, recent SNe Ia and supernova remnants, or in the event of a detection, provide the first measurement of the luminosity and temperature of a SN Ia progenitor.

\section*{Acknowledgements}
I.R.S. acknowledges support from the Australian Research Council Future Fellowship Grant FT160100028.

%%%%%%%%%%%%%%%%%%%%%%%%
\bibliographystyle{mnras}
\bibliography{viitteet}

%%%%%%%%%%%%%%%%%%%%%%%%
\bsp	% typesetting comment
\label{lastpage}
\end{document}